# Multi-Gateway Cooperation in Multibeam Satellite Systems

Gan Zheng*, Symeon Chatzinotas*, Björn Ottersten*†
*SnT - securityandtrust.lu, University of Luxembourg,
Email: {Gan.Zheng, Symeon.Chatzinotas, Bjorn.Ottersten}@uni.lu
†Royal Institute of Technology (KTH), Sweden, Email: bjorn.ottersten@ee.kth.se

*Abstract*—Multibeam systems with hundreds of beams have been recently deployed in order to provide higher capacities by employing fractional frequency reuse. Furthermore, employing full frequency reuse and precoding over multiple beams has shown great throughput potential in literature. However, feeding all this data from a single gateway is not feasible based on the current frequency allocations. In this context, we investigate a range of scenarios involving beam clusters where each cluster is managed by a single gateway. More specifically, the following cases are considered for handling intercluster interference: a) conventional frequency colouring, b) joint processing within cluster, c) partial CSI sharing among clusters, d) partial CSI and data sharing among clusters. CSI sharing does not provide considerable performance gains with respect to b) but combined with data sharing offers roughly a 40% improvement over a) and a 15% over b).

## I. INTRODUCTION

In existing literature, the majority of satellite studies in the area of multiuser precoding and detection assume an ideal situation, where a single GateWay (GW) has access to data streams and channel state information (CSI) for all system users. However in reality the single GW assumption is unrealistic as it would imply a large feeder link bandwidth which is not available under current frequency allocations. A proposed solution would be to move feeder links into the Q/V bands, but this requires expensive site diversity to tackle outages due to atmospheric effects. A more feasible solution which is currently used (e.g. by KASat) is to deploy a large number of GWs in order to reuse feeder link frequencies through directive antennas[1]. In a conventional colouring scheme, multiple GWs do not seriously affect the transmission techniques since each beam is still individually processed, but under full frequency reuse multiple GWs create intrasystem interference and limit the potential of multibeam processing. In this context, this paper studies and compares the performance of conventional and multibeam processing of multiple GWs considering different levels of available CSI and data sharing.

Some additional arguments for the deployment of multiple GWs include reduced multibeam processing complexity, redundancy in case of failure and shorter distribution network loops. In more detail, having multiple GWs means that each GW has to handle a smaller number of beams which implies reduced signal processing complexity. In addition, in case one of the GWs fail the traffic can be rerouted through other GWs to avoid service outage. Finally, content distributors (e.g. media companies, data centers) are geographically distributed and thus multiple GWs could reduce the amount of backhauling to the GW.

The remainder of this paper is structured as follows. Section II overviews related work mainly from both terrestrial and satellite literature, while section III presents the system model. Subsequently, section IV investigates non-cooperative multiGW techniques while section V proposes and analyzes cooperative multiGW techniques. Section VI evaluates and compares the aforementioned techniques based on numerical results. A promising cluster overlapping is introduced in Section VII as a future direction and Section VIII concludes this paper.

## II. RELATED WORK

This section overviews related results on multiuser joint precoding from satellite literature, followed by some base station cooperation techniques applied in a terrestrial context.

### A. Satellite Literature

Multiuser precoding for satellite communications was considered in [1] for a wide range of performance objectives (sum-rate, rate balancing, rate matching etc) and for generic linear and non-linear power constraints. Furthermore, the effect of power sharing among beams was investigated in [2] for a rate balancing problem under linear beamforming. Finally, the energy efficiency of MMSE beamforming was studied in [3]. The reader is referred to [1] for a more detailed review of multibeam precoding.

### B. Terrestrial Literature

In terrestrial wireless communications, several multi-cell processing methods for the forward link were devised in [4]–[6]. In particular, assuming data sharing, [4] studied the design of transmit beamforming by recasting the downlink beamforming problem into a least minimum mean-square-error (LMMSE) estimation problem. [5] proposed a distributed design in time-division-duplex (TDD) systems using only local CSI and data sharing and demonstrated that performance close to the Pareto bound can be obtained. Recently, [6] proposed distributed multicell processing without data or CSI sharing but required moderate control signaling among BSs.

---

[1]It should be noted that even with highly directive antennas the feeder links originating in different GWs are partially interfering. However, in this work we assume that multiple GWs are sufficiently separated on the earth surface so that feeder link interference can be ignored.

## III. SYSTEM MODEL

We consider a multibeam system with $C$ GWs each serving a cluster of $K$ beams using $K$ antenna feeds. The overall bandwidth is $W$. Our aim is to design effective transmission schemes to serve the whole $CK$ beams over the coverage area using multibeam processing and GW cooperation. For simplicity, we assume that beams managed by a single GW are subject to a total power constraint $P_T$.

The channel from the $c_1$-th GW's onboard antenna feeds to the $k$-th user in the $c_2$-th cluster is denoted by $\mathbf{h}_{c1,c2,k}$. In the following, we drop the indices and give the details about channel modeling between any satellite feed and the terminal antenna.

### A. Channel Model

The employed channel model refers to fixed mobile services where rain fading is the main impairment. In addition, the power level received at each terminal depends on the beam pattern and its position within the coverage area. More specifically, the channel vector for the $k$th user can be written as:

$$\mathbf{h}_k = \varsigma_k \boldsymbol{\xi}_k \odot \mathbf{b}_k^{\frac{1}{2}} \qquad (1)$$

where $\boldsymbol{\xi}$, $\mathbf{b}$ and $\varsigma_k$ denote the rain fading, beam gain and path loss respectively. The rain fading coefficients from all satellite feeds (independently of the operating GW) towards a single terminal antenna are given in the following vector

$$\boldsymbol{\xi}_k = \xi_k^{-\frac{1}{2}} e^{-\mathrm{j}\phi_k} \mathbf{1}_N \qquad (2)$$

where $\phi$ denotes a uniformly distributed phase. The rain attenuation coefficients $\xi_k$ are modeled according to ITU-R[2] Recommendation P.1853 [7] with parameters listed in Table I. The path loss is calculated based on the employed frequency and the slant range towards the satellite which depends on the terminal position:

$$\varsigma_k = (4\pi\lambda d_k)^2 \qquad (3)$$

where $d_k$ is the slant range for the $k$th user. Then the beam gain can be approximated by [8]:

$$b(\theta, k) = b_{\max}(k) \left( \frac{J_1(u)}{2u} + 36 \frac{J_3(u)}{u^3} \right)^2 \qquad (4)$$

where $\theta$ is the angle between the beam center and the terminal location and $\theta_{3\mathrm{dB}}$ is the angle which corresponds to 3-dB power loss. In addition, the auxiliary variable $u$ is defined as $u = 2.07123 \sin\theta / \sin\theta_{3\mathrm{dB}}$ and $J_1$ and $J_3$ are the first-kind Bessel function of order 1 and 3. The coefficient $b_{\max}(k)$ represents the gain at the $k$th beam center.

[2]International Telecommunications Union - Radiocommunications Sector.

TABLE I
SATELLITE SCENARIO PARAMETERS

| Parameter | Value |
|---|---|
| Orbit | GEO |
| Frequency band | 20 GHz |
| Number of beams | $K = 133$ |
| Coverage area diameter | $D = 500$ km |
| 3dB angle | $\theta_{3\mathrm{dB}} = 0.4^o$ |
| Rain fading mean | $\mu = -3.4249$ |
| Rain fading variance | $\sigma = 1.5768$ |
| Polarization | Single |
| Max antenna Tx gain | 52dBi |
| User terminal maximum antenna gain | 41.7dBi |
| FL free space loss | 210dB |
| User link bandwidth | W=500MHz |
| Clear sky receiver temperature | $207^o$K |

## IV. NON-COOPERATIVE TRANSMISSION STRATEGIES

We first review the transmission schemes without using GW cooperation. Each GW is responsible for serving its own users within its cluster by performing either single-beam of multibeam processing, which will result in causing uncontrolled interference to and receiving interference from the other clusters as well.

### A. Signal Model

Suppose the data intended for user $k$ in the $c$-th cluster sent from its serving GW is $s_{ck}$ with $\mathbb{E}[|s_{ck}|^2] = 1$. Before transmitted, it is weighted by the beamforming vector $\sqrt{p_{ck}}\mathbf{w}_{ck}$ where $\|\mathbf{w}_{ck}\| = 1$ and $p_{ck}$ represents the transmit power. The received signal at user $k$ in the $c$-th cell is

$$y_{ck} = \mathbf{h}_{c,c,k}^\dagger \left( \sum_{k=1}^K \sqrt{p_{ck}} \mathbf{w}_{ck} s_{ck} + n_k \right), \qquad (5)$$

where $n_k$ is the independent and identically distributed (i.i.d.) zero-mean Gaussian random noise with power density $N_0$. The achievable rate is given by

$$R_{ck} = \log(1 + \Gamma_{ck}), \qquad (6)$$

Then the received SINR at user $k$ in the $c$-th cluster is

$$\Gamma_{ck} = \frac{p_{ck}|\mathbf{w}_{ck}^\dagger \mathbf{h}_{c,c,k}|^2}{\sum_{\{b,j\} \neq \{c,k\}} p_{bj}|\mathbf{w}_{bj}^\dagger \mathbf{h}_{b,c,k}|^2 + WN_0}. \qquad (7)$$

### B. Conventional Coloring Scheme

In this scheme, each beam is served independently of each other and interference is mitigated by using frequency reuse plan (4-colour scheme ). As a result, employing multiple GWs does affect sum-rate performance since the serving GWs needs only access to data and CSI for the desired beam. In this case, the rate is given by

$$R_{ck} = \frac{1}{4} \log(1 + \Gamma_{ck}), \qquad (8)$$

and the received SINR at user $k$ can be written as

$$\Gamma_{ck} = \frac{p_{ck}|h_{c,c,k}|^2}{\sum_b |h_{b,c,k}|^2 + 4WN_0}, \qquad (9)$$

where $b$ is the index of a cluster using the same frequency color as cluster $c$.

## C. Joint Multibeam Processing Within a Cluster Only

If there is a single GW in the system covering the whole area, assuming full frequency reuse, it has been shown that multibeam processing can achieve significant performance gain over the above coloring scheme [1].

In the context of multiple GWs, since there is no coordination among them, each GW processes its beams independently. Consequently, interference among clusters has to be tolerated and there is no interference mitigation among clusters.

In this paper, we adopt throughput as the design objective. In this case, the $c$-th GW tries to maximize its own cluster's throughput, i.e.,

$$\max_{\mathbf{w}_1,p_1,\cdots,\mathbf{w}_K,p_K} \sum_{k=1}^{K} \underline{R_{ck}} \quad (10)$$
$$\text{s.t.} \quad \sum_{k=1}^{K} p_{ck} \leq P_T.$$

Note that the $c$-th GW does not have channel knowledge outside its cluster, the rate is expressed as

$$\underline{R_{ck}} \triangleq \log\left(1 + \frac{p_{ck}|\mathbf{w}_{ck}^\dagger \mathbf{h}_{c,c,k}|^2}{\sum_{j\neq k} p_{cj}|\mathbf{w}_{cj}^\dagger \mathbf{h}_{c,c,k}|^2 + WN_0}\right). \quad (11)$$

which is not achievable because inter-cluster interference is not considered. In our simulation, we will evaluate the performance of this design considering the interference from other clusters. In the following, we briefly describe how to design the beamforming and power.

To achieve a good balance between complexity and performance, Regularized Zero-Forcing (R-ZF) precoding was proposed where a regularization parameter is introduced to take into account of the noise effect which is ignored in ZF precoding. To be specific, the beamforming vector $\mathbf{w}_{ck}$ is taken from the normalized $k$-th column of

$$\mathbf{W}_c = \left(\mathbf{H}_c^\dagger \mathbf{H}_c + \beta \mathbf{I}\right)^{-1} \mathbf{H}_c^\dagger, \quad (12)$$

where $\mathbf{H}_c$ collects all users' vector channels and $\beta$ is the regularization factor, which needs to be carefully chosen to achieve good performance. Based on the large system analysis, the optimal $\beta$ (in the statistical sense) to maximize the SINR is given by [9],

$$\beta^{opt} = \frac{N_0 WK}{P_T}. \quad (13)$$

With R-ZF precoding, problems (10) reduce to the following power allocation problem:

$$\max_{p_{ck}} \sum_{k=1}^{K} \log\left(1 + \frac{p_{ck}|\mathbf{w}_{ck}^\dagger \mathbf{h}_{c,c,k}|^2}{\sum_{j\neq k} p_{cj}|\mathbf{w}_{cj}^\dagger \mathbf{h}_{c,c,k}|^2 + WN_0}\right)$$
$$\text{s.t.} \quad \sum_{k=1}^{K} p_{ck} \leq P_T \quad (14)$$

and locally convergent numerical algorithms can be applied to find the suboptimal power allocation.

## V. PROPOSED COOPERATIVE CLUSTER TRANSMISSION STRATEGIES

### A. Signal Model

In this case, we assume that each user can be served by more than one GW. Suppose the data intended for user $k$ in the $c$-th cluster, $s_{ck}$ with $\mathbb{E}[|s_{ck}|^2] = 1$, is transmit from GWs in the set $\mathcal{A}_{ck}$. Obviously its own serving GW $c \in \mathcal{A}_{ck}$. Similarly, the $c$-th GW transmits to the set of users $\mathcal{G}_c$. It can be easily seen that $ck \in \mathcal{G}_c$ if and only if $c \in \mathcal{A}_{ck}$.

For a cooperating GW $b \in \mathcal{A}_{ck}$, the allocated transmit power for $s_{ck}$ is $p_{bck}$ and before transmitted, it is weighted by the beamforming vector $\sqrt{p_{bck}}\mathbf{w}_{bck}$ where $\|\mathbf{w}_{bck}\| = 1$. At the terminal side, the received SINR is

$$\Gamma_{ck} = \frac{|\sum_{b \in \mathcal{A}_{ck}} \sqrt{p_{bck}} \mathbf{w}_{bck}^\dagger \mathbf{h}_{b,c,k}|^2}{\sum_{\{b,j\}\neq\{c,k\}} |\sum_{d \in \mathcal{A}_{bj}} \sqrt{p_{dbj}} \mathbf{w}_{dbj}^\dagger \mathbf{h}_{d,c,k}|^2 + WN_0}. \quad (15)$$

### B. Joint Hyper-Cluster Processing with partial CSI Sharing

We assume that a set of neighboring clusters form a hyper-cluster to mitigate interference. The cooperation is on the CSI level rather than data level and therefore $|\mathcal{G}_c| = K$. Suppose the $c$-th GW can see interference to neighboring users in a set $\mathcal{B}_c$ therefore has CSI to those users. We then optimize beamforming vectors to maximize the signal-to-leakage-and-noise ratio (SLNR) below for user $k$ in cluster $c$,

$$\text{SLNR}_{cck} = \frac{|\mathbf{w}^\dagger \mathbf{h}_{c,c,k}|^2}{\sum_{j\neq k}|\mathbf{w}^\dagger \mathbf{h}_{c,c,j}|^2 + \sum_{(b,j)\in\mathcal{B}_c}|\mathbf{w}^\dagger \mathbf{h}_{c,b,j}|^2 + WN_0}. \quad (16)$$

Then the beamforming design is given by (17). With this beamforming, power optimization in cluster $c$ is similar to (14) to maximize each cluster's own throughput and can be written as

$$\max_{\{p_{cck}\}} \sum_{k=1}^{K} \log\left(1 + \frac{|\sqrt{p_{cck}}\mathbf{w}_{cck}^\dagger \mathbf{h}_{c,c,k}|^2}{\sum_{j\neq k}|\sqrt{p_{ccj}}\mathbf{w}_{ccj}^\dagger \mathbf{h}_{c,c,k}|^2 + WN_0}\right)$$
$$\text{s.t.} \quad \sum_{k=1}^{K} p_{ck} \leq P_T. \quad (18)$$

Note that the achievable rate for user $k$ in cluster $c$ is evaluated as

$$R_{ck} = \log\left(1 + \frac{|\sqrt{p_{cck}}\mathbf{w}_{cck}^\dagger \mathbf{h}_{c,c,k}|^2}{\sum_{\{b,j\}\neq\{c,k\}}|\sqrt{p_{bbj}}\mathbf{w}_{bbj}^\dagger \mathbf{h}_{b,c,k}|^2 + WN_0}\right). \quad (19)$$

### C. Joint Hyper-Cluster Processing with partial CSI and partial data sharing

Data sharing entails that a cluster has access to the user data destined for users at the edge of adjacent clusters. This creates an opportunity for helping instead of interfering with cluster-edge users. More specifically, a GW can evaluate the channel gains towards users in adjacent beams and select/schedule the user(s) which would receive the strongest interference based on the channel norm. Assuming that data sharing is enabled,

$$\mathbf{w}_{cck} = \arg\min_{\|\mathbf{w}\|=1} \mathrm{SLNR}_{cck}$$

$$= \frac{\left(\sum_{j\neq k}\mathbf{h}_{c,c,j}\mathbf{h}^{\dagger}_{c,c,j} + \sum_{(b,j)\in\mathcal{B}_c}\mathbf{h}_{c,b,j}\mathbf{h}^{\dagger}_{c,b,j} + WN_0\mathbf{I}\right)^{-1}\mathbf{h}_{c,c,k}}{\left\|\left(\sum_{j\neq k}\mathbf{h}_{c,c,j}\mathbf{h}^{\dagger}_{c,c,j} + \sum_{(b,j)\in\mathcal{B}_c}\mathbf{h}_{c,b,j}\mathbf{h}^{\dagger}_{c,b,j} + WN_0\mathbf{I}\right)^{-1}\mathbf{h}_{c,c,k}\right\|}. \quad (17)$$

the GW can include this user in its precoding set and transmit a useful signal towards this user. It should be noted that the user will be also served by its own cluster/GW so it will benefit from a power gain.[3] Ideally, more that one users could be selected/schedule but this would generally result in increased multiuser interference because the transmit dimensions (cluster feeds) are fewer than the receive dimensions (scheduled users). Taking this observation into account, we consider that $M << K$ additional users can be scheduled in each cluster and and therefore $|\mathcal{G}_c| = K + M$. The additional users are the users with the strongest channel norm $|\mathbf{h}_{c,b,j}|^2$ towards the $c$-th GW. It should be noted that CSI is also available for all users in $\mathcal{G}_c$, namely $\mathcal{G}_c \subseteq \mathcal{B}_c$.

We then optimize beamforming vectors for all users in $\mathcal{G}_c$ to maximize the signal-to-leakage-and-noise ratio (SLNR), which is shown in (20).

Then the beamforming design is achieved by (21). With this beamforming, power optimization in cluster $c$ is similar to (14) to maximize each cluster's own throughput and can be written as

$$\max_{\{p_{(g,l)\in\mathcal{G}_c}\}} \sum_{k=1}^{K+M} \log\left(1 + \frac{|\sqrt{p_{cgl}}\mathbf{w}^{\dagger}_{cgl}\mathbf{h}_{c,g,l}|^2}{\sum_{\substack{(g,j)\in\mathcal{G}_c\\j\neq l}}|\sqrt{p_{cgj}}\mathbf{w}^{\dagger}_{cgj}\mathbf{h}_{c,g,l}|^2 + WN_0}\right)$$

$$\text{s.t.} \sum_{k=1}^{K+M} p_{gl} \leq P_T. \quad (22)$$

Note that the achievable rate for user $(g,l) \in \mathcal{G}_c$ is evaluated as

$$R_{gl} = \log\left(1 + \frac{|\sqrt{p_{cgl}}\mathbf{w}^{\dagger}_{cgl}\mathbf{h}_{c,g,l}|^2}{\sum_{\{b,j\}\neq\{g,l\}}|\sqrt{p_{bbj}}\mathbf{w}^{\dagger}_{bbj}\mathbf{h}_{b,g,l}|^2 + WN_0}\right). \quad (23)$$

## VI. NUMERICAL RESULTS

Simulation results are provided to compare the performance of the proposed hyper-cluster scheme with the single cluster multibeam processing and a conventional spotbeam system. We consider a system with $C = 19$ clusters, each having $K = 7$ antenna feeds and serving $K = 7$ beams, so in total 133 beams are being served, as depicted in Fig. 1. For the proposed GW cooperation, the following 7 hyper-clusters are formed with $C_1 = \{3, 9, 10\}, C_2 = \{4, 11, 12\}, C_3 = \{2, 8, 19\}, C_4 = \{5, 13, 14\}, C_5 = \{7, 17, 18\}, C_6 = \{6, 15, 16\}, C_7 = \{1\}$, where each set contains the indices of GWs in that hyper-cluster.

[3]assuming that the two transmission are synchronized and can be coherently decoded.

For the proposed GW cooperation, we assume in a hyper-cluster, each GW selects one user ($M = 1$) with strongest channel strength in each of its neighboring cluster, therefore obtains CSI to those users and serves those users as well (for the case of data sharing).

We choose the average per beam throughput as the performance metric. In Fig. 2, the achievable per beam throughputs are shown for different schemes against per beam power. It is observed that the conventional 4-coloring scheme has the lowest achievable rate due to the lack of multibeam processing. The individual cluster multibeam processing without GW coordination achieves reasonable performance and the proposed multi-GW scheme with CSI sharing shows marginal performance gain and this is due to the lack of enough transmit dimensions to suppress interference. Using extra data sharing among non-overlapping GWs gives about 15% higher throughput than the one with only CSI.

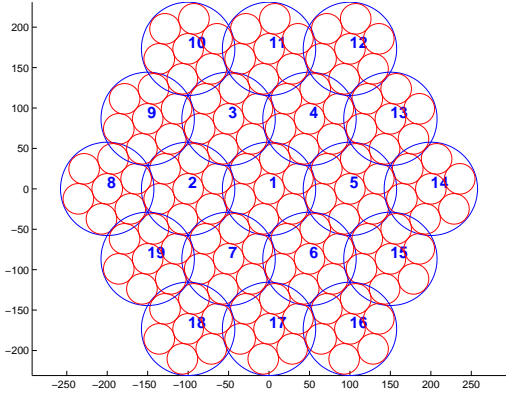

Fig. 1. Clusters of Beams.

## VII. FUTURE WORK

### A. Joint Hyper-Cluster Processing with Overlapping Clusters

According to this technique, a subset of beams at the edge of adjacent clusters are served by multiple clusters by sharing user data and onboard antenna feeds. This can be also seen as having partially overlapping clusters. Users in the overlapped area receive the sum of multiple signals from cooperating GWs which have to be properly designed according to the performance objective. It should also be noted that overlapping clusters create interdependencies among the power constraints of clusters (due to sharing of antenna feeds) and a limited coordination scheme may be needed to ensure that they are not violated. As described before, each GW has access

$$\text{SLNR}_{(g,l)\in\mathcal{G}_c} = \frac{|\mathbf{w}^\dagger \mathbf{h}_{c,g,l}|^2}{\sum_{\substack{(g,j)\in\mathcal{G}_c \\ j\neq l}} |\mathbf{w}^\dagger \mathbf{h}_{c,g,j}|^2 + \sum_{(b,j)\in\mathcal{B}_c} |\mathbf{w}^\dagger \mathbf{h}_{c,b,j}|^2 + WN_0}, \quad \forall l = 1\ldots K+M. \tag{20}$$

$$\begin{aligned}\mathbf{w}_{(g,l)\in\mathcal{G}_c} &= \arg\min_{\|\mathbf{w}\|=1} \text{SLNR}_{(g,l)\in\mathcal{G}_c} \\ &= \frac{\left(\sum_{\substack{(g,j)\in\mathcal{G}_c \\ j\neq l}} \mathbf{h}_{c,g,j}\mathbf{h}_{c,g,j}^\dagger + \sum_{(b,j)\in\mathcal{B}_c} \mathbf{h}_{c,b,j}\mathbf{h}_{c,b,j}^\dagger + WN_0\mathbf{I}\right)^{-1} \mathbf{h}_{c,g,l}}{\left\|\left(\sum_{\substack{(g,j)\in\mathcal{G}_c \\ j\neq l}} \mathbf{h}_{c,g,j}\mathbf{h}_{c,g,j}^\dagger + \sum_{(b,j)\in\mathcal{B}_c} \mathbf{h}_{c,b,j}\mathbf{h}_{c,b,j}^\dagger + WN_0\mathbf{I}\right)^{-1} \mathbf{h}_{c,g,l}\right\|}.\end{aligned} \tag{21}$$

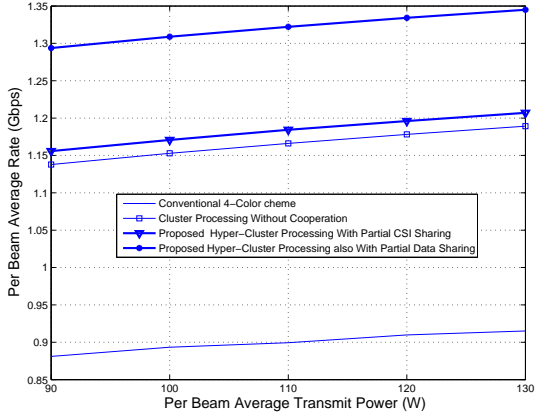

Fig. 2. Comparison of achievable per beam throughput.

to partial CSI, namely channels towards its own users and towards interfering users in adjacent clusters. The SINR can be expressed as:

$$\Gamma_{ck} = \frac{|\sqrt{p_{cck}}\mathbf{w}_{cck}^\dagger \mathbf{h}_{c,c,k}|^2}{\sum_{\{b,j\}\neq\{c,k\}} |\sqrt{p_{bbj}}\mathbf{w}_{bbj}^\dagger \mathbf{h}_{b,c,k}|^2 + WN_0}. \tag{24}$$

for the users in the cluster center and

$$\Gamma_{ck} = \frac{|\sum_{b\in\mathcal{A}_{ck}} \sqrt{p_{bck}}\mathbf{w}_{bck}^\dagger \mathbf{h}_{b,c,k}|^2}{\sum_{\{b,j\}\neq\{c,k\}} |\sum_{d\in\mathcal{A}_{bj}} \sqrt{p_{dbj}}\mathbf{w}_{dbj}^\dagger \mathbf{h}_{d,c,k}|^2 + WN_0}, \tag{25}$$

for the users in the overlapped clusters. It should be noted that since the clusters are partially overlapping, the channel vectors $\mathbf{h}_{b,c,k}$ will have a number of common elements for beam indices. The optimization algorithm is going to exploit this fact in order to design precoders which optimize the performance objective. In addition, these are some options for modeling the power constraints:

1) Individual power constrains $P$ for non overlapping beams and $P/K$ for beams that can be accessed by $K$ clusters (realistic and simple)
2) Individual power constrains $P$ for all beams and cluster coordination to ensure that overlapping beams transmit within limits (most realistic but most complicated)
3) Sum power constrains for all cluster beams and cluster coordination to ensure that clusters transmit within limits (less realistic and complicated)

## VIII. CONCLUSIONS

In this paper, we have considered the problem of multi-GW cooperation in the forward link of multibeam satellite systems with fixed service. Specifically, we have proposed two cooperation schemes: one is to share partial CSI and the other assumes partial data can be shared on top of CSI sharing. We have compared the per beam throughput achieved by the proposed schemes with non-cooperative strategies including conventional 4-coloring schemes and single GW multibeam processing. Due to the obvious lack of more antenna feeds than the served users, the proposed scheme with CSI sharing does not give significant gain compared to the single GW multibeam processing while the proposed data sharing indeed greatly outperforms the non-cooperative solutions. We also point out a promising direction worth of future study, where cluster overlapping is allowed, and the performance could be improved at the cost of extra signaling between GWs.

## IX. ACKNOWLEDGMENT

This work was supported by the National Research Fund, Luxembourg under the CORE project "CO$^2$SAT: Cooperative and Cognitive Architectures for Satellite Networks".